\documentclass[12pt,a4paper]{article}
\usepackage[english]{babel}
\usepackage{graphicx}
\usepackage[authoryear,round]{natbib}
\usepackage{hyperref}

\begin{document}
\title{\bf Catastrophic events in protoplanetary disks and their observational manifestations}
\author{Tatiana V. Demidova$^1$ and Vladimir P. Grinin$^{2,3}$}
\date{}
\maketitle
\noindent$^1$ --- Crimean Astrophysical Observatory, Nauchny, Crimea, Russia \\ 
email: {\tt proxima1@list.ru}\\ 
$^2$ --- Pulkovo Observatory of the Russian Academy of Sciences, St. Petersburg, Russia\\
$^3$ --- V.V. Sobolev Astronomical Institute, St. Petersburg University, St. Petersburg, Russia
\begin{abstract}
Observations of protoplanetary disks with high angular resolution
using an ALMA interferometer showed that ring-shaped structures
are often visible in their images, indicating strong disturbances
in the disks. The mechanisms of their formation are vividly
discussed in the literature. This article shows that the formation
of such structures can be the result of destructive collisions of
large bodies (planetesimals and planetary embryos) accompanied by
the formation of a large number of dust particles, and the
subsequent evolution of a cloud of dust formed in this way.
\end{abstract}

\section{Introduction} \label{sec:intro}
One of the most interesting results obtained using ALMA and VLT is the detection of substructures on images of
protoplanetary disks observed with high angular resolution \citep[see, e.g.][]{2018ApJ...863...44A, 2018A&A...619A.171B,
2018A&A...619A.161C, 2018ApJ...869L..43H, 2018ApJ...869L..50P,
2017ApJ...837..132V}. Statistical analysis shows
\citep{2018ApJ...869L..43H} that the observed substructures often have ring-like form, less often they look like wrapped spiral arms and very rarely exhibit
crescent-like azimuthal asymmetries. The rings usually have fuzzy
(blurred) boundaries. The origin of such structures is most often
associated with disturbances that arise when planets interact with
a protoplanetary disk \citep[e.g.,][]{2013A&A...549A..97R, 2015A&A...579A.106V, 2015ApJ...809...93D, 2017ApJ...843..127D, 2018ApJ...866..110D, 2016MNRAS.463L..22D, 2016ApJ...818...76J, 2017ApJ...850..201B}.
However, since no direct observational signs of the existence of
planets in these disks were found, alternative models have also been
considered~\citep[e.g.,][]{2015ApJ...815L..15B, 2015ApJ...806L...7Z, 2015ApJ...813L..14B, 2016ApJ...821...82O}. Most of these models associate the formation of substructures
with the development of various kinds of instabilities in
disks~\citep[e.g.,][]{2009ApJ...697.1269J, 2014ApJ...796...31B, 2014ApJ...794...55T, 2018A&A...609A..50D}.

Below we show that the formation of observable structures can be
the result of a collision of large bodies occurring in the disks
during the formation of planetary systems. Such collisions play an
important role in enlarging the embryos of the planets, from which
planets are then formed~\citep{1998Icar..136..304C, 2006AJ....131.1837K, 2016ApJ...825...33K}. But not all collisions lead to the union
of colliding bodies. In cases where the velocity of their relative
motion is large enough, the collision process ends with the
destruction of bodies. As a result, many smaller bodies and
particles are formed. Such processes are cascade in nature, and
they are described in the literature~\citep[see, e.g.,][]{2012MNRAS.425..657J, 2015ApJ...810..136G}. There is
a lot of evidence that they occurred during the birth of
the Solar system. In particular, the widespread hypotheses about
the origin of the Moon connects its formation with the destructive
collision of the proto-Earth with a large
body~\citep{1976LPI.....7..120C}. The formation of the Mercury and origin of the Pluto--Charon system is associated  with large impacts \citep{2007SSRv..132..189B, 2005Sci...307..546C, 2006Natur.439..946S, 2011AJ....141...35C}. It was suggested that the large obliquity of Uranus is also connected with the giant impact \citep[e.g.][]{1992Icar...99..167S}.

It is obvious that the collisions of such large bodies, which
result in the formation of a dust cloud, should be accompanied by
flashes in the infrared region of the spectrum. In this
connection, the results of ~\citet{2014Sci...345.1032M} are
of great interest. They observed strong changes in the IR
luminosity of a young star and interpreted them as the result of
collisions of large bodies in a circumstellar disk~\citep{2015ApJ...805...77M, 2019AJ....157..202S}. 

In paper~\citet{2014MNRAS.440.3757J} was shown that the asymmetries seen at large separations in some debris disks, like Beta Pictoris, could be the result of violent impacts. The authors showed that after the impact
the debris forms an expanding cloud that follows the progenitor
object. The rate of the cloud shears depends  on the
velocity dispersion of the debris. But the dispersion of the cloud will typically take no more than one orbit. Then the initial debris clump has been sheared out into a spiral structure. The shearing process will continue until the spiral has been completely merged to a ring during several orbits. The ring structure may be saved for $10^4$ convolutions. The authors also showed that millimeter-size and
larger grains will follow these distributions, while smaller grains will be blown-out by the radiation. We consider a similar model of disintegration of the dust cloud but in the presence of gas.

\section{Initial condition}
We assumed a model that consists of a young star of solar mass ($ M_{\ast} =
M_{\odot} $) embedded in a gas disk with total mass is $ M_{disk}
= 0.01M_{\odot} $. At the beginning of simulation the disk matter was distributed to be azimuthally symmetrical within the radii $ r_{in} = 0.2$ and
$r_{out} = 100$ AU. The initial density distribution of the disk
is
\begin{equation}
\rho(r,z)=\frac{\Sigma_0}{\sqrt{2\pi}H(r)}\Big(\frac{r}{r_{in}}\Big)^{-1.5}e^{-\frac{z^2}{2H^2(r)}},
\end{equation}
where $\Sigma_0$ is arbitrary scale parameter, which is determined by
disk mass. Hydrostatic scale height is $H(r)=\sqrt{\frac{\kappa T_{mid}(r) r^3}{GM_{\ast} \mu m_H}}$, where $\kappa$, $G$ and $m_H$ are
the Boltzmann constant, the gravitational constant and the mass of
a hydrogen atom. $\mu=2.35$ is the mean molecular weight
\citep{1994A&A...286..149D}. Following~\citet{1997ApJ...490..368C}
we determine the law of midplane temperature distribution
$T_{mid}(r)=\sqrt[4]{\frac{\phi}{4}}\sqrt{\frac{R_{\ast}}{r}}T_{\ast}$,
where $\phi=0.05$ \citep{2004A&A...421.1075D}. It was assumed that the disk is isothermal in the vertical direction along $z$. The
temperature of the star was assumed to be $T_{\ast} = 5780$ K and
the star radius was $R_{\ast}=R_{\odot}$.

A dust cloud (the result of the collision) was placed at a distance of  $30$ AU from the star. Its mass was $0.02\%$ (approximately equal to the half of the Moon mass) of the mass of the disk's dust component. It was supposed that the dust-to-gas mass ratio was $1/100$ ($M_{dust} = 10^{-4}M_{\odot}$). We simulate the cloud matter consisting of a number of particle types with different sizes from $1$~$\mu$m to $1$~mm with the density of dust particles equal to $\rho_{d} = 1$~g~cm$^{-3}$. The particles of the cloud were distributed according to the law
$\rho_w(x)\propto e^{-x}$, where $x$ is the dimensionless distance
from the center of the cloud, and the characteristic size
of the cloud was $0.3H(r=30)\approx 0.48$ AU.

It was assumed that the dust cloud expands slowly during the process of orbital
motion. So at the time $t = 0$ the velocity of motion of particles in the cloud is equal to the local Keplerian velocity at a given point of the disk plus random velocities of particles. The latter were distributed over a Gaussian with a mathematical expectation of $V=50$~m~s$^{-1}$ and a standard deviation of $50$~m~s$^{-1}$.

\begin{figure*}[ht!]
\includegraphics[width=0.85\columnwidth]{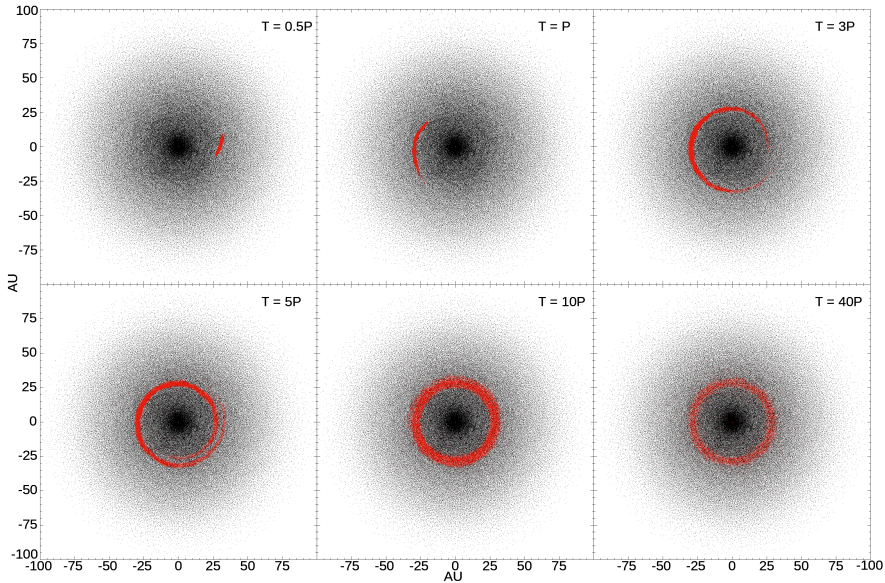}
\caption{\normalsize Particle distribution in the model. Gas
particles are shown in black and dust particles are red dots. The
time in the upper left corner of the images is indicated in the
orbital periods of the cloud center ($P$). \label{fig:Part}}
\end{figure*}

\section{Model rationale}
As stated in Sec.~\ref{sec:intro} the formation of the planetary system may be accompanied with catastrophic collisions of planet embryos or planetesimals. In paper~\citet{2019AJ....157..202S} was shown the formation of cloud of small dust may take days to a month depending on the velocity of impact. But the conclusions regarding the cascade destruction of colliding bodies are made for the models without gas; however, they are applicable to our model. The fact is that at the periphery of the disk the gas density is low in order to significantly slow down the movement of debris. The density and temperature of the gas at $30$ AU are $\rho(30,0)=7.5\cdot 10^{-14}$~g~cm$^{-3}$ and $T(30)=24$ K, the mean free path of a molecule is $\lambda=2.4\cdot 10^5$~cm, and the speed of sound is $c_s=2.9\cdot 10^4$~cm~s$^{-1}$. Therefore the stopping time is $t_e=\frac{s\rho_d}{c_s\rho}$  for dust particles with size $s<\frac{9}{4}\lambda$ \citep[Epstein regime in][]{1977MNRAS.180...57W}. For particles with $s=1$~mm this time is $1.53$ yr and it increases with $s$. Even this value is significantly more than the destruction time obtained in~\citet{2019AJ....157..202S}.

The critical energy (at which the target body loses at least half of its mass) was estimated in dependence on the radius of the body \citep[e.g.,][]{2015Icar..262...58G}. This dependence was extended to the case of planetary embryos by~\citet{2009ApJ...700L.118M}, which allowed us to estimate the critical velocity of a head-on collision of two bodies of quarter lunar mass $V_i\approx 5.42$ km~s$^{-1}$. This value is approximately equal to the velocity of the Keplerian motion at a distance of 30 AU ($V_k\approx 5.44$ km~s$^{-1}$). But $V_i$ increases with decreasing mass of the projectile body. 

However, the initial mass of colliding planetesimals can be several times smaller than that accepted in our model. After the cascade process, the fragments' velocity of expansion should be larger than the two-body surface escape velocity ($v_{esc}$). Assume that the total mass of colliding planetesimals was equal to $10\%$ ($m_{d}=3.68\cdot10^{24}$~g) of that accepted in the model, and the initial cloud expansion velocity was $V_0=2.06v_{esc}\approx1.47$~km~s$^{-1}$ ($v_{esc}\approx 0.72$~km~s$^{-1}$). The velocity of dust particle size of $1$ mm decreased according to the law $V(t)=V_0 e^{-t/te} $ due to the gas drag forces to a value $V=50$~m~s$^{-1}$ during $T\approx5.16$~yrs. The cloud expanded and its size reached $\Delta r\approx0.48$~AU. Since planetesimals are present in the disk, we assume that $90\%$ of dust matter is a large particles that settled down to the disk plane. The fragments of the cloud exchange of the momentum with the disk particles, which in turn become a part of the cloud. Since the surface density of the dust at $30$~AU is $\Sigma_d (30)=0.045$~g~cm$^{-2}$ the cloud particles can interact with dusty matter with mass 

$
M_{dd}=0.9 \Sigma_d (30)\int\limits_0^T \int\limits_0^{\Delta  r}\frac{V_k dt}{2r}[(r+dr)^2-(r-dr)^2]dr=\\=0.9 \Sigma_d (30) 2 V_k V_0\int\limits_0^T[\int\limits_0^t e^{-t'/te} dt']dt\approx 3.31\cdot 10^{25}~g
$\\
orbiting over time $T$. The sum of the masses $M_{dd}+m_d=3.68\cdot10^{25}$~g is equal to the mass accepted in our model. In that case the critical velocity of impact is $V_i\approx 2.52$~km~s$^{-1}$. However, the probability of collision of such planetesimals in the absence of external influence is negligible. Therefore, a mechanism is needed to disperse the planetary embryo to high speed.

\begin{figure*}[ht!]
\includegraphics[width=0.9\columnwidth]{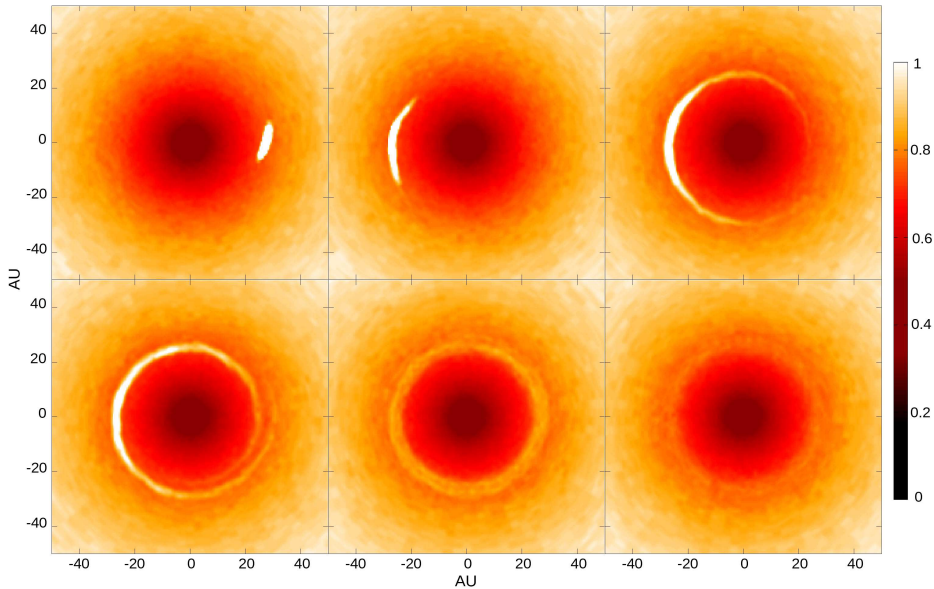}
\caption{\normalsize Disk images at wavelength $1.3$ mm. The color
shows the intensity of radiation multiplied by $R^2 $ in arbitrary
units. The time points are in accordance with
Fig.~\ref{fig:Part}.\label{fig:Img}}
\end{figure*}

We suppose that a giant planet may be responsible for the scattering of solid bodies to the outer part of the disk at high speeds. The eccentricity and inclination of particles orbits will be damping due to resonant interactions with a gas disk on the secular time scale \citep[see, e.g.,][]{2012ARA&A..50..211K}. But the orbit of planetary embryos may cross the planetary chaotic zone during the migration of one or several planets. In \cite{2013A&A...556A..28B} was shown the semi-major axis and eccentricity of the particles may change significantly on an orbital timescale under the condition of a rapid dynamical chaos. This mechanism can contribute to the transfer of planetary embryos to the periphery of the disk. Some of them leave their initial position at high velocity~\citep{2015ApJ...799...41M}. In~\citet{2008ApJS..179..451K} was shown the process of formation of planetary embryos in the outer part of the protoplanetary disk proceeds more slowly than in the inner one. However, several hundred $\sim 100$ km planetesimals can exist on the periphery of the disk and may collide with scattering planetary embryos.

\section{Simulation}
We calculated the evolution of the cloud in gas surroundings using
the SPH method (smooth particle hydrodynamics). The calculations
of the dust particle dynamics in the gaseous medium were performed
using the code Gadget-2~\citep{2001NewA....6...79S,
2005MNRAS.364.1105S} modified in \citet{2016Ap.....59..449D}.
Equations describing the interaction of dust and gas from the
work~\citet{2014MNRAS.440.2147L} were added to the code. In total,
$2\cdot 10^5$ particles with gas properties and $5\cdot 10^4$ with
dust properties took part in the simulation.

The dust opacity is calculated using Mie theory for magnesium-iron
silicates~\citep{1995A&A...300..503D}. The RADMC-3D
code was used for the 3D radiative transfer calculations. It is
the open code which available online at \url{http://www.ita.uni-heidelberg.de/~dullemond/software/radmc-3d/}.

\section{Results and Discussion}
The results of the hydrodynamic calculations are
presented on Fig.~\ref{fig:Part}. We see that,
due to the differential rotation of the Keplerian disk, the dust
cloud stretches and successively takes all three forms described
above: a piece of arc resembling a cyclonic vortex and a tightly
wound spiral, which then turns into a ring. 

To calculate the observable images of the disk it was assumed that
$10\%$ of dust mass is in small grains ($0.1$~$\mu$m) which
determine the disk temperature structure and such particles
well mixed with the gas matter. The images of the disk on the
wavelength $1.3$~mm are presented on~Fig.~\ref{fig:Img}. Naturally the particles of $1$ mm give the greatest contribution to the luminosity of the structure. Whereas particles of $10$~$\mu$m and less are quickly dispersed by gas and can be considered well mixed with gas.

The contrast of the structures formed by cloud matter is
determined by the contribution of small dust to the background
thermal radiation of the disk. The structures at the initial
stages of cloud disintegration are visible also in the case
when $100\%$ of the matter of the dust part of the disk is small
dust. However, at a later stages, the contrast of the structure
decreases considerably. 

According to Fig.~\ref{fig:Part} the transition from a crescent-like
structure to a ring occurs during several revolutions. The ring
exists much longer, during several dozen revolutions. Its borders
are gradually diffused with a time. Calculations have
shown that the speed of transition from the cloud to the ring depends on the velocity of dust cloud expansion. The process may last a dozen revolutions in the model with zero expanding velocity. Immediately after the collisions the embryos' fragment velocities will be high (comparable to the collision speeds). They are diminished in cascade collisions and by the interaction with the protoplanetary disk. As a result, the fragments become smaller and slow down their movement to the exchange of momentum with the surrounding matter, which becomes a part of the primary cloud. This means that the initial mass of colliding bodies may be significantly less than that adopted in our model, and accordingly, the collision velocity should be much smaller.

The advantage of our model is that it is based on a physical process (catastrophic collisions of the massive bodies) which observational manifestations have
evidential facts in the history of the Solar system. Such destructive collisions can probably be initiated by one or several giant planets in the process of their migration. As mentioned in Sec.~\ref{sec:intro} a giant planet itself can form a ring-shaped structure in a protoplanetary disk, and a giant impact produce another one. Thus, the considered model contributes to the explanation of the multiple rings in the protoplanetary disk images.
 
We assume that the formation of giant planets may be accompanied by several destructive collisions of planetary embryos. To confirm this hypothesis, numerical calculations on secular time-scales are required, as well as taking the gas presence into account. If it is confirmed by further researches, this will allow for the observation of  the consequences of catastrophic collisions at the birth of other
planetary systems by methods of interferometry.

{\bf Acknowledgments}. It is a pleasure to thank the referee for valuable and useful remarks. T.V.D. is supported by the Russian Science Foundation under grant 19-72-10063. V.P.G. is supported by the Program of Fundamental Research of the Russian Academy of Sciences under grant KP19-270. 



\bibliographystyle{unsrtnat}
\bibliography{biblio}

\begin{thebibliography}{52}
\providecommand{\natexlab}[1]{#1}
\providecommand{\url}[1]{\texttt{#1}}
\expandafter\ifx\csname urlstyle\endcsname\relax
  \providecommand{\doi}[1]{doi: #1}\else
  \providecommand{\doi}{doi: \begingroup \urlstyle{rm}\Url}\fi

\bibitem[{Avenhaus} et~al.(2018){Avenhaus}, {Quanz}, {Garufi}, {Perez},
  {Casassus}, {Pinte}, {Bertrang}, {Caceres}, {Benisty}, and
  {Dominik}]{2018ApJ...863...44A}
Henning {Avenhaus}, Sascha~P. {Quanz}, Antonio {Garufi}, Sebastian {Perez},
  Simon {Casassus}, Christophe {Pinte}, Gesa H.~M. {Bertrang}, Claudio
  {Caceres}, Myriam {Benisty}, and Carsten {Dominik}.
\newblock {Disks around T Tauri Stars with SPHERE (DARTTS-S). I. SPHERE/IRDIS
  Polarimetric Imaging of Eight Prominent T Tauri Disks}.
\newblock \emph{Astrophys. J.}, 863\penalty0 (1):\penalty0 44, Aug 2018.
\newblock \doi{10.3847/1538-4357/aab846}.

\bibitem[{Benisty} et~al.(2018){Benisty}, {Juh{\'a}sz}, {Facchini}, {Pinilla},
  {de Boer}, {P{\'e}rez}, {Keppler}, {Muro-Arena}, {Villenave}, {Andrews},
  {Dominik}, {Dullemond}, {Gallenne}, {Garufi}, {Ginski}, and
  {Isella}]{2018A&A...619A.171B}
M.~{Benisty}, A.~{Juh{\'a}sz}, S.~{Facchini}, P.~{Pinilla}, J.~{de Boer}, L.~M.
  {P{\'e}rez}, M.~{Keppler}, G.~{Muro-Arena}, M.~{Villenave}, S.~{Andrews},
  C.~{Dominik}, C.~P. {Dullemond}, A.~{Gallenne}, A.~{Garufi}, C.~{Ginski}, and
  A.~{Isella}.
\newblock {Shadows and asymmetries in the T Tauri disk HD 143006: evidence for
  a misaligned inner disk}.
\newblock \emph{Astron. Astrophys.}, 619:\penalty0 A171, Nov 2018.
\newblock \doi{10.1051/0004-6361/201833913}.

\bibitem[{Cazzoletti} et~al.(2018){Cazzoletti}, {van Dishoeck}, {Pinilla},
  {Tazzari}, {Facchini}, {van der Marel}, {Benisty}, {Garufi}, and
  {P{\'e}rez}]{2018A&A...619A.161C}
P.~{Cazzoletti}, E.~F. {van Dishoeck}, P.~{Pinilla}, M.~{Tazzari},
  S.~{Facchini}, N.~{van der Marel}, M.~{Benisty}, A.~{Garufi}, and L.~M.
  {P{\'e}rez}.
\newblock {Evidence for a massive dust-trapping vortex connected to spirals.
  Multi-wavelength analysis of the HD 135344B protoplanetary disk}.
\newblock \emph{Astron. Astrophys.}, 619:\penalty0 A161, Nov 2018.
\newblock \doi{10.1051/0004-6361/201834006}.

\bibitem[{Huang} et~al.(2018){Huang}, {Andrews}, {P{\'e}rez}, {Zhu},
  {Dullemond}, {Isella}, {Benisty}, {Bai}, {Birnstiel}, {Carpenter},
  {Guzm{\'a}n}, {Hughes}, {{\"O}berg}, {Ricci}, {Wilner}, and
  {Zhang}]{2018ApJ...869L..43H}
Jane {Huang}, Sean~M. {Andrews}, Laura~M. {P{\'e}rez}, Zhaohuan {Zhu},
  Cornelis~P. {Dullemond}, Andrea {Isella}, Myriam {Benisty}, Xue-Ning {Bai},
  Tilman {Birnstiel}, John~M. {Carpenter}, Viviana~V. {Guzm{\'a}n}, A.~Meredith
  {Hughes}, Karin~I. {{\"O}berg}, Luca {Ricci}, David~J. {Wilner}, and Shangjia
  {Zhang}.
\newblock {The Disk Substructures at High Angular Resolution Project (DSHARP).
  III. Spiral Structures in the Millimeter Continuum of the Elias 27, IM Lup,
  and WaOph 6 Disks}.
\newblock \emph{Astrophys. J. Lett.}, 869\penalty0 (2):\penalty0 L43, Dec 2018.
\newblock \doi{10.3847/2041-8213/aaf7a0}.

\bibitem[{P{\'e}rez} et~al.(2018){P{\'e}rez}, {Benisty}, {Andrews}, {Isella},
  {Dullemond}, {Huang}, {Kurtovic}, {Guzm{\'a}n}, {Zhu}, {Birnstiel}, {Zhang},
  {Carpenter}, {Wilner}, {Ricci}, {Bai}, {Weaver}, and
  {{\"O}berg}]{2018ApJ...869L..50P}
Laura~M. {P{\'e}rez}, Myriam {Benisty}, Sean~M. {Andrews}, Andrea {Isella},
  Cornelis~P. {Dullemond}, Jane {Huang}, Nicol{\'a}s~T. {Kurtovic}, Viviana~V.
  {Guzm{\'a}n}, Zhaohuan {Zhu}, Tilman {Birnstiel}, Shangjia {Zhang}, John~M.
  {Carpenter}, David~J. {Wilner}, Luca {Ricci}, Xue-Ning {Bai}, Erik {Weaver},
  and Karin~I. {{\"O}berg}.
\newblock {The Disk Substructures at High Angular Resolution Project (DSHARP).
  X. Multiple Rings, a Misaligned Inner Disk, and a Bright Arc in the Disk
  around the T Tauri star HD 143006}.
\newblock \emph{Astrophys. J. Lett.}, 869\penalty0 (2):\penalty0 L50, Dec 2018.
\newblock \doi{10.3847/2041-8213/aaf745}.

\bibitem[{van Boekel} et~al.(2017){van Boekel}, {Henning}, {Menu}, {de Boer},
  {Langlois}, {M{\"u}ller}, {Avenhaus}, {Boccaletti}, {Schmid}, {Thalmann},
  {Benisty}, {Dominik}, {Ginski}, {Girard}, {Gisler}, {Lobo Gomes}, {Menard},
  {Min}, {Pavlov}, {Pohl}, {Quanz}, {Rabou}, {Roelfsema}, {Sauvage}, {Teague},
  {Wildi}, and {Zurlo}]{2017ApJ...837..132V}
R.~{van Boekel}, Th. {Henning}, J.~{Menu}, J.~{de Boer}, M.~{Langlois},
  A.~{M{\"u}ller}, H.~{Avenhaus}, A.~{Boccaletti}, H.~M. {Schmid}, Ch.
  {Thalmann}, M.~{Benisty}, C.~{Dominik}, Ch. {Ginski}, J.~H. {Girard},
  D.~{Gisler}, A.~{Lobo Gomes}, F.~{Menard}, M.~{Min}, A.~{Pavlov}, A.~{Pohl},
  S.~P. {Quanz}, P.~{Rabou}, R.~{Roelfsema}, J.~F. {Sauvage}, R.~{Teague},
  F.~{Wildi}, and A.~{Zurlo}.
\newblock {Three Radial Gaps in the Disk of TW Hydrae Imaged with SPHERE}.
\newblock \emph{Astrophys. J.}, 837\penalty0 (2):\penalty0 132, Mar 2017.
\newblock \doi{10.3847/1538-4357/aa5d68}.

\bibitem[{Ruge} et~al.(2013){Ruge}, {Wolf}, {Uribe}, and
  {Klahr}]{2013A&A...549A..97R}
J.~P. {Ruge}, S.~{Wolf}, A.~L. {Uribe}, and H.~H. {Klahr}.
\newblock {Tracing large-scale structures in circumstellar disks with ALMA}.
\newblock \emph{Astron. Astrophys.}, 549:\penalty0 A97, Jan 2013.
\newblock \doi{10.1051/0004-6361/201220390}.

\bibitem[{van der Marel} et~al.(2015){van der Marel}, {van Dishoeck},
  {Bruderer}, {P{\'e}rez}, and {Isella}]{2015A&A...579A.106V}
N.~{van der Marel}, E.~F. {van Dishoeck}, S.~{Bruderer}, L.~{P{\'e}rez}, and
  A.~{Isella}.
\newblock {Gas density drops inside dust cavities of transitional disks around
  young stars observed with ALMA}.
\newblock \emph{Astron. Astrophys.}, 579:\penalty0 A106, Jul 2015.
\newblock \doi{10.1051/0004-6361/201525658}.

\bibitem[{Dong} et~al.(2015){Dong}, {Zhu}, and {Whitney}]{2015ApJ...809...93D}
Ruobing {Dong}, Zhaohuan {Zhu}, and Barbara {Whitney}.
\newblock {Observational Signatures of Planets in Protoplanetary Disks I. Gaps
  Opened by Single and Multiple Young Planets in Disks}.
\newblock \emph{Astrophys. J.}, 809\penalty0 (1):\penalty0 93, Aug 2015.
\newblock \doi{10.1088/0004-637X/809/1/93}.

\bibitem[{Dong} et~al.(2017){Dong}, {Li}, {Chiang}, and
  {Li}]{2017ApJ...843..127D}
Ruobing {Dong}, Shengtai {Li}, Eugene {Chiang}, and Hui {Li}.
\newblock {Multiple Disk Gaps and Rings Generated by a Single Super-Earth}.
\newblock \emph{Astrophys. J.}, 843\penalty0 (2):\penalty0 127, Jul 2017.
\newblock \doi{10.3847/1538-4357/aa72f2}.

\bibitem[{Dong} et~al.(2018){Dong}, {Li}, {Chiang}, and
  {Li}]{2018ApJ...866..110D}
Ruobing {Dong}, Shengtai {Li}, Eugene {Chiang}, and Hui {Li}.
\newblock {Multiple Disk Gaps and Rings Generated by a Single Super-Earth. II.
  Spacings, Depths, and Number of Gaps, with Application to Real Systems}.
\newblock \emph{Astrophys. J.}, 866\penalty0 (2):\penalty0 110, Oct 2018.
\newblock \doi{10.3847/1538-4357/aadadd}.

\bibitem[{Demidova} and {Shevchenko}(2016)]{2016MNRAS.463L..22D}
Tatiana~V. {Demidova} and Ivan~I. {Shevchenko}.
\newblock {Three-lane and multilane signatures of planets in planetesimal
  discs}.
\newblock \emph{MNRAS}, 463\penalty0 (1):\penalty0 L22--L25, Nov 2016.
\newblock \doi{10.1093/mnrasl/slw150}.

\bibitem[{Jin} et~al.(2016){Jin}, {Li}, {Isella}, {Li}, and
  {Ji}]{2016ApJ...818...76J}
Sheng {Jin}, Shengtai {Li}, Andrea {Isella}, Hui {Li}, and Jianghui {Ji}.
\newblock {Modeling Dust Emission of HL Tau Disk Based on Planet-Disk
  Interactions}.
\newblock \emph{Astrophys. J.}, 818\penalty0 (1):\penalty0 76, Feb 2016.
\newblock \doi{10.3847/0004-637X/818/1/76}.

\bibitem[{Bae} et~al.(2017){Bae}, {Zhu}, and {Hartmann}]{2017ApJ...850..201B}
Jaehan {Bae}, Zhaohuan {Zhu}, and Lee {Hartmann}.
\newblock {On the Formation of Multiple Concentric Rings and Gaps in
  Protoplanetary Disks}.
\newblock \emph{Astrophys. J.}, 850\penalty0 (2):\penalty0 201, Dec 2017.
\newblock \doi{10.3847/1538-4357/aa9705}.

\bibitem[{Banzatti} et~al.(2015){Banzatti}, {Pinilla}, {Ricci}, {Pontoppidan},
  {Birnstiel}, and {Ciesla}]{2015ApJ...815L..15B}
A.~{Banzatti}, P.~{Pinilla}, L.~{Ricci}, K.~M. {Pontoppidan}, T.~{Birnstiel},
  and F.~{Ciesla}.
\newblock {Direct Imaging of the Water Snow Line at the Time of Planet
  Formation using Two ALMA Continuum Bands}.
\newblock \emph{Astrophys. J. Lett.}, 815\penalty0 (1):\penalty0 L15, Dec 2015.
\newblock \doi{10.1088/2041-8205/815/1/L15}.

\bibitem[{Zhang} et~al.(2015){Zhang}, {Blake}, and
  {Bergin}]{2015ApJ...806L...7Z}
Ke~{Zhang}, Geoffrey~A. {Blake}, and Edwin~A. {Bergin}.
\newblock {Evidence of Fast Pebble Growth Near Condensation Fronts in the HL
  Tau Protoplanetary Disk}.
\newblock \emph{Astrophys. J. Lett.}, 806\penalty0 (1):\penalty0 L7, Jun 2015.
\newblock \doi{10.1088/2041-8205/806/1/L7}.

\bibitem[{Birnstiel} et~al.(2015){Birnstiel}, {Andrews}, {Pinilla}, and
  {Kama}]{2015ApJ...813L..14B}
Tilman {Birnstiel}, Sean~M. {Andrews}, Paola {Pinilla}, and Mihkel {Kama}.
\newblock {Dust Evolution Can Produce Scattered Light Gaps in Protoplanetary
  Disks}.
\newblock \emph{Astrophys. J.l}, 813\penalty0 (1):\penalty0 L14, Nov 2015.
\newblock \doi{10.1088/2041-8205/813/1/L14}.

\bibitem[{Okuzumi} et~al.(2016){Okuzumi}, {Momose}, {Sirono}, {Kobayashi}, and
  {Tanaka}]{2016ApJ...821...82O}
Satoshi {Okuzumi}, Munetake {Momose}, Sin-iti {Sirono}, Hiroshi {Kobayashi},
  and Hidekazu {Tanaka}.
\newblock {Sintering-induced Dust Ring Formation in Protoplanetary Disks:
  Application to the HL Tau Disk}.
\newblock \emph{Astrophys. J.}, 821\penalty0 (2):\penalty0 82, Apr 2016.
\newblock \doi{10.3847/0004-637X/821/2/82}.

\bibitem[{Johansen} et~al.(2009){Johansen}, {Youdin}, and
  {Klahr}]{2009ApJ...697.1269J}
A.~{Johansen}, A.~{Youdin}, and H.~{Klahr}.
\newblock {Zonal Flows and Long-lived Axisymmetric Pressure Bumps in
  Magnetorotational Turbulence}.
\newblock \emph{Astrophys. J.}, 697\penalty0 (2):\penalty0 1269--1289, Jun
  2009.
\newblock \doi{10.1088/0004-637X/697/2/1269}.

\bibitem[{Bai} and {Stone}(2014)]{2014ApJ...796...31B}
Xue-Ning {Bai} and James~M. {Stone}.
\newblock {Magnetic Flux Concentration and Zonal Flows in Magnetorotational
  Instability Turbulence}.
\newblock \emph{Astrophys. J.}, 796\penalty0 (1):\penalty0 31, Nov 2014.
\newblock \doi{10.1088/0004-637X/796/1/31}.

\bibitem[{Takahashi} and {Inutsuka}(2014)]{2014ApJ...794...55T}
Sanemichi~Z. {Takahashi} and Shu-ichiro {Inutsuka}.
\newblock {Two-component Secular Gravitational Instability in a Protoplanetary
  Disk: A Possible Mechanism for Creating Ring-like Structures}.
\newblock \emph{Astrophys. J.}, 794\penalty0 (1):\penalty0 55, Oct 2014.
\newblock \doi{10.1088/0004-637X/794/1/55}.

\bibitem[{Dullemond} and {Penzlin}(2018)]{2018A&A...609A..50D}
C.~P. {Dullemond} and A.~B.~T. {Penzlin}.
\newblock {Dust-driven viscous ring-instability in protoplanetary disks}.
\newblock \emph{Astron. Astrophys.}, 609:\penalty0 A50, Jan 2018.
\newblock \doi{10.1051/0004-6361/201731878}.

\bibitem[{Chambers} and {Wetherill}(1998)]{1998Icar..136..304C}
J.~E. {Chambers} and G.~W. {Wetherill}.
\newblock {Making the Terrestrial Planets: N-Body Integrations of Planetary
  Embryos in Three Dimensions}.
\newblock \emph{Icarus}, 136\penalty0 (2):\penalty0 304--327, Dec 1998.
\newblock \doi{10.1006/icar.1998.6007}.

\bibitem[{Kenyon} and {Bromley}(2006)]{2006AJ....131.1837K}
Scott~J. {Kenyon} and Benjamin~C. {Bromley}.
\newblock {Terrestrial Planet Formation. I. The Transition from Oligarchic
  Growth to Chaotic Growth}.
\newblock \emph{Astron. J.}, 131\penalty0 (3):\penalty0 1837--1850, Mar 2006.
\newblock \doi{10.1086/499807}.

\bibitem[{Kenyon} and {Bromley}(2016)]{2016ApJ...825...33K}
Scott~J. {Kenyon} and Benjamin~C. {Bromley}.
\newblock {Making Planet Nine: Pebble Accretion at 250-750 AU in a
  Gravitationally Unstable Ring}.
\newblock \emph{Astrophys. J.}, 825\penalty0 (1):\penalty0 33, Jul 2016.
\newblock \doi{10.3847/0004-637X/825/1/33}.

\bibitem[{Jackson} and {Wyatt}(2012)]{2012MNRAS.425..657J}
Alan~P. {Jackson} and Mark~C. {Wyatt}.
\newblock {Debris from terrestrial planet formation: the Moon-forming
  collision}.
\newblock \emph{MNRAS}, 425\penalty0 (1):\penalty0 657--679, Sep 2012.
\newblock \doi{10.1111/j.1365-2966.2012.21546.x}.

\bibitem[{Genda} et~al.(2015{\natexlab{a}}){Genda}, {Kobayashi}, and
  {Kokubo}]{2015ApJ...810..136G}
H.~{Genda}, H.~{Kobayashi}, and E.~{Kokubo}.
\newblock {Warm Debris Disks Produced by Giant Impacts during Terrestrial
  Planet Formation}.
\newblock \emph{Astrophys. J.}, 810\penalty0 (2):\penalty0 136, Sep
  2015{\natexlab{a}}.
\newblock \doi{10.1088/0004-637X/810/2/136}.

\bibitem[{Cameron} and {Ward}(1976)]{1976LPI.....7..120C}
A.~G.~W. {Cameron} and W.~R. {Ward}.
\newblock {The Origin of the Moon}.
\newblock In \emph{Lunar and Planetary Science Conference}, volume~7, page 120,
  Mar 1976.

\bibitem[{Benz} et~al.(2007){Benz}, {Anic}, {Horner}, and
  {Whitby}]{2007SSRv..132..189B}
W.~{Benz}, A.~{Anic}, J.~{Horner}, and J.~A. {Whitby}.
\newblock {The Origin of Mercury}.
\newblock \emph{Space Science Reviews}, 132\penalty0 (2-4):\penalty0 189--202,
  Oct 2007.
\newblock \doi{10.1007/s11214-007-9284-1}.

\bibitem[{Canup}(2005)]{2005Sci...307..546C}
Robin~M. {Canup}.
\newblock {A Giant Impact Origin of Pluto-Charon}.
\newblock \emph{Science}, 307\penalty0 (5709):\penalty0 546--550, Jan 2005.
\newblock \doi{10.1126/science.1106818}.

\bibitem[{Stern} et~al.(2006){Stern}, {Weaver}, {Steffl}, {Mutchler},
  {Merline}, {Buie}, {Young}, {Young}, and {Spencer}]{2006Natur.439..946S}
S.~A. {Stern}, H.~A. {Weaver}, A.~J. {Steffl}, M.~J. {Mutchler}, W.~J.
  {Merline}, M.~W. {Buie}, E.~F. {Young}, L.~A. {Young}, and J.~R. {Spencer}.
\newblock {A giant impact origin for Pluto's small moons and satellite
  multiplicity in the Kuiper belt}.
\newblock \emph{Nature}, 439\penalty0 (7079):\penalty0 946--948, Feb 2006.
\newblock \doi{10.1038/nature04548}.

\bibitem[{Canup}(2011)]{2011AJ....141...35C}
Robin~M. {Canup}.
\newblock {On a Giant Impact Origin of Charon, Nix, and Hydra}.
\newblock \emph{Astron. J.}, 141\penalty0 (2):\penalty0 35, Feb 2011.
\newblock \doi{10.1088/0004-6256/141/2/35}.

\bibitem[{Slattery} et~al.(1992){Slattery}, {Benz}, and
  {Cameron}]{1992Icar...99..167S}
Wayne~L. {Slattery}, Willy {Benz}, and A.~G.~W. {Cameron}.
\newblock {Giant impacts on a primitive Uranus}.
\newblock \emph{Icarus}, 99\penalty0 (1):\penalty0 167--174, Sep 1992.
\newblock \doi{10.1016/0019-1035(92)90180-F}.

\bibitem[{Meng} et~al.(2014){Meng}, {Su}, {Rieke}, {Stevenson}, {Plavchan},
  {Rujopakarn}, {Lisse}, {Poshyachinda}, and {Reichart}]{2014Sci...345.1032M}
Huan Y.~A. {Meng}, Kate Y.~L. {Su}, George~H. {Rieke}, David~J. {Stevenson},
  Peter {Plavchan}, Wiphu {Rujopakarn}, Carey~M. {Lisse}, Saran {Poshyachinda},
  and Daniel~E. {Reichart}.
\newblock {Large impacts around a solar-analog star in the era of terrestrial
  planet formation}.
\newblock \emph{Science}, 345\penalty0 (6200):\penalty0 1032--1035, Aug 2014.
\newblock \doi{10.1126/science.1255153}.

\bibitem[{Meng} et~al.(2015){Meng}, {Su}, {Rieke}, {Rujopakarn}, {Myers},
  {Cook}, {Erdelyi}, {Maloney}, {McMath}, {Persha}, {Poshyachinda}, and
  {Reichart}]{2015ApJ...805...77M}
Huan Y.~A. {Meng}, Kate Y.~L. {Su}, George~H. {Rieke}, Wiphu {Rujopakarn},
  Gordon {Myers}, Michael {Cook}, Emery {Erdelyi}, Chris {Maloney}, James
  {McMath}, Gerald {Persha}, Saran {Poshyachinda}, and Daniel~E. {Reichart}.
\newblock {Planetary Collisions Outside the Solar System: Time Domain
  Characterization of Extreme Debris Disks}.
\newblock \emph{Astrophys. J.}, 805\penalty0 (1):\penalty0 77, May 2015.
\newblock \doi{10.1088/0004-637X/805/1/77}.

\bibitem[{Su} et~al.(2019){Su}, {Jackson}, {G{\'a}sp{\'a}r}, {Rieke}, {Dong},
  {Olofsson}, {Kennedy}, {Leinhardt}, {Malhotra}, {Hammer}, {Meng},
  {Rujopakarn}, {Rodriguez}, {Pepper}, {Reichart}, {James}, and
  {Stassun}]{2019AJ....157..202S}
Kate Y.~L. {Su}, Alan~P. {Jackson}, Andr{\'a}s {G{\'a}sp{\'a}r}, George~H.
  {Rieke}, Ruobing {Dong}, Johan {Olofsson}, G.~M. {Kennedy}, Zo{\"e}~M.
  {Leinhardt}, Renu {Malhotra}, Michael {Hammer}, Huan Y.~A. {Meng},
  W.~{Rujopakarn}, Joseph~E. {Rodriguez}, Joshua {Pepper}, D.~E. {Reichart},
  David {James}, and Keivan~G. {Stassun}.
\newblock {Extreme Debris Disk Variability: Exploring the Diverse Outcomes of
  Large Asteroid Impacts During the Era of Terrestrial Planet Formation}.
\newblock \emph{Astron. J.}, 157\penalty0 (5):\penalty0 202, May 2019.
\newblock \doi{10.3847/1538-3881/ab1260}.

\bibitem[{Jackson} et~al.(2014){Jackson}, {Wyatt}, {Bonsor}, and
  {Veras}]{2014MNRAS.440.3757J}
Alan~P. {Jackson}, Mark~C. {Wyatt}, Amy {Bonsor}, and Dimitri {Veras}.
\newblock {Debris froms giant impacts between planetary embryos at large
  orbital radii}.
\newblock \emph{MNRAS}, 440\penalty0 (4):\penalty0 3757--3777, Jun 2014.
\newblock \doi{10.1093/mnras/stu476}.

\bibitem[{Dutrey} et~al.(1994){Dutrey}, {Guilloteau}, and
  {Simon}]{1994A&A...286..149D}
A.~{Dutrey}, S.~{Guilloteau}, and M.~{Simon}.
\newblock {Images of the GG Tauri rotating ring}.
\newblock \emph{Astron. Astrophys.}, 286:\penalty0 149--159, June 1994.

\bibitem[{Chiang} and {Goldreich}(1997)]{1997ApJ...490..368C}
E.~I. {Chiang} and P.~{Goldreich}.
\newblock {Spectral Energy Distributions of T Tauri Stars with Passive
  Circumstellar Disks}.
\newblock \emph{Astrophys. J.}, 490:\penalty0 368--376, November 1997.
\newblock \doi{10.1086/304869}.

\bibitem[{Dullemond} and {Dominik}(2004)]{2004A&A...421.1075D}
C.~P. {Dullemond} and C.~{Dominik}.
\newblock {The effect of dust settling on the appearance of protoplanetary
  disks}.
\newblock \emph{Astron. Astrophys.}, 421:\penalty0 1075--1086, July 2004.
\newblock \doi{10.1051/0004-6361:20040284}.

\bibitem[{Weidenschilling}(1977)]{1977MNRAS.180...57W}
S.~J. {Weidenschilling}.
\newblock {Aerodynamics of solid bodies in the solar nebula.}
\newblock \emph{MNRAS}, 180:\penalty0 57--70, Jul 1977.
\newblock \doi{10.1093/mnras/180.1.57}.

\bibitem[{Genda} et~al.(2015{\natexlab{b}}){Genda}, {Fujita}, {Kobayashi},
  {Tanaka}, and {Abe}]{2015Icar..262...58G}
Hidenori {Genda}, Tomoaki {Fujita}, Hiroshi {Kobayashi}, Hidekazu {Tanaka}, and
  Yutaka {Abe}.
\newblock {Resolution dependence of disruptive collisions between planetesimals
  in the gravity regime}.
\newblock \emph{Icarus}, 262:\penalty0 58--66, Dec 2015{\natexlab{b}}.
\newblock \doi{10.1016/j.icarus.2015.08.029}.

\bibitem[{Marcus} et~al.(2009){Marcus}, {Stewart}, {Sasselov}, and
  {Hernquist}]{2009ApJ...700L.118M}
Robert~A. {Marcus}, Sarah~T. {Stewart}, Dimitar {Sasselov}, and Lars
  {Hernquist}.
\newblock {Collisional Stripping and Disruption of Super-Earths}.
\newblock \emph{Astrophys. J. Lett.}, 700\penalty0 (2):\penalty0 L118--L122,
  Aug 2009.
\newblock \doi{10.1088/0004-637X/700/2/L118}.

\bibitem[{Kley} and {Nelson}(2012)]{2012ARA&A..50..211K}
W.~{Kley} and R.~P. {Nelson}.
\newblock {Planet-Disk Interaction and Orbital Evolution}.
\newblock \emph{Ann. Rev. Astron. Astrophys.}, 50:\penalty0 211--249, Sep 2012.
\newblock \doi{10.1146/annurev-astro-081811-125523}.

\bibitem[{Batygin} and {Morbidelli}(2013)]{2013A&A...556A..28B}
K.~{Batygin} and A.~{Morbidelli}.
\newblock {Analytical treatment of planetary resonances}.
\newblock \emph{Astron. Astrophys.}, 556:\penalty0 A28, Aug 2013.
\newblock \doi{10.1051/0004-6361/201220907}.

\bibitem[{Morrison} and {Malhotra}(2015)]{2015ApJ...799...41M}
Sarah {Morrison} and Renu {Malhotra}.
\newblock {Planetary Chaotic Zone Clearing: Destinations and Timescales}.
\newblock \emph{Astrophys. J.}, 799\penalty0 (1):\penalty0 41, Jan 2015.
\newblock \doi{10.1088/0004-637X/799/1/41}.

\bibitem[{Kenyon} and {Bromley}(2008)]{2008ApJS..179..451K}
Scott~J. {Kenyon} and Benjamin~C. {Bromley}.
\newblock {Variations on Debris Disks: Icy Planet Formation at 30-150 AU for
  1-3 M$_{☉}$ Main-Sequence Stars}.
\newblock \emph{Astrophys. J. Supp.}, 179\penalty0 (2):\penalty0 451--483, Dec
  2008.
\newblock \doi{10.1086/591794}.

\bibitem[{Springel} et~al.(2001){Springel}, {Yoshida}, and
  {White}]{2001NewA....6...79S}
Volker {Springel}, Naoki {Yoshida}, and Simon D.~M. {White}.
\newblock {GADGET: a code for collisionless and gasdynamical cosmological
  simulations}.
\newblock \emph{New Astronomy}, 6\penalty0 (2):\penalty0 79--117, Apr 2001.
\newblock \doi{10.1016/S1384-1076(01)00042-2}.

\bibitem[{Springel}(2005)]{2005MNRAS.364.1105S}
Volker {Springel}.
\newblock {The cosmological simulation code GADGET-2}.
\newblock \emph{MNRAS}, 364\penalty0 (4):\penalty0 1105--1134, Dec 2005.
\newblock \doi{10.1111/j.1365-2966.2005.09655.x}.

\bibitem[{Demidova}(2016)]{2016Ap.....59..449D}
T.~V. {Demidova}.
\newblock {Modelling the Gas Dynamics of Protoplanetary Disks by the SPH
  Method}.
\newblock \emph{Astrophysics}, 59\penalty0 (4):\penalty0 449--460, Dec 2016.
\newblock \doi{10.1007/s10511-016-9448-3}.

\bibitem[{Laibe} and {Price}(2014)]{2014MNRAS.440.2147L}
Guillaume {Laibe} and Daniel~J. {Price}.
\newblock {Dusty gas with one fluid in smoothed particle hydrodynamics}.
\newblock \emph{MNRAS}, 440\penalty0 (3):\penalty0 2147--2163, May 2014.
\newblock \doi{10.1093/mnras/stu359}.

\bibitem[{Dorschner} et~al.(1995){Dorschner}, {Begemann}, {Henning}, {Jaeger},
  and {Mutschke}]{1995A&A...300..503D}
J.~{Dorschner}, B.~{Begemann}, T.~{Henning}, C.~{Jaeger}, and H.~{Mutschke}.
\newblock {Steps toward interstellar silicate mineralogy. II. Study of
  Mg-Fe-silicate glasses of variable composition.}
\newblock \emph{Astron. Astrophys.}, 300:\penalty0 503, Aug 1995.

\end{thebibliography}



\end{document}